\documentclass[varenna]{cimento}

\usepackage{graphicx}  

\begin{document}

\title{Strangeness in Relativistic Astrophysics}
 
\author{J\"urgen Schaffner--Bielich}

\institute{
Institut f\"ur Theoretische Physik/Astrophysik,
J. W. Goethe Universit\"at,
Max-von-Laue-Str.~1,
D-60438 Frankfurt/Main, Germany}

\author{Stefan Schramm}

\institute{
Center for Scientific Computing and
Institut f\"ur Theoretische Physik/Astrophysik,
J. W. Goethe Universit\"at,
Max-von-Laue-Str.~1,
D-60438 Frankfurt/Main, Germany}

\author{Horst St\"ocker}

\institute{
Gesellschaft f\"ur Schwerionenforschung,
D-64291 Darmstadt, Germany\\
Frankfurt Institute for Advanced Studies (FIAS),
J. W. Goethe Universit\"at,
Ruth-Moufang-Str.~1,
D-60438 Frankfurt/Main, Germany\\
Institut f\"ur Theoretische Physik/Astrophysik,
J. W. Goethe Universit\"at,
Max-von-Laue-Str.~1,
D-60438 Frankfurt/Main, Germany}

\shortauthor{J. Schaffner--Bielich, S. Schramm, \atque H. St\"ocker}

\PACSes{
\PACSit{26.60.+c}{Nuclear matter aspects of neutron stars}
\PACSit{97.60.Jd}{Neutron stars}
\PACSit{21.80.+a}{Hypernuclei}
\PACSit{12.38.Mh}{Quark-gluon plasma}
} 

\maketitle

\begin{abstract}
  In these lecture notes, the role of strangeness in relativistic
  astrophysics of compact stars is addressed. The appearance of strange
  particles, as hyperons, kaons, and strange quarks, in the
  core of compact stars is examined and common features as well as
  differences are presented. Impacts on the global properties of compact
  stars and signals of the presence of exotic matter are outlined for
  the various strange phases which can appear in the interior at high
  densities.
\end{abstract}


\section{Introduction}

Neutron stars are the final endpoint in the evolution of stars more
massive than eight solar masses. They are born in a spectacular
explosive event, a core-collapse supernova, which can outshine an entire
galaxy in its brightness. A surprisingly good estimate about the
characteristic masses and radii can be made by adopting the argument
of Landau \cite{Landau32}. The delicate balance between gravitational
energy and Fermi energy can only be achieved up to a certain critical
number of nucleons (fermions) which is $N\approx (M_P/m_N)^3 \approx
10^{57}$, where $M_P$ stands for the Planck mass and $m_N$ for the
nucleon mass. The corresponding maximum mass for neutron stars amounts
to $M_{\rm max} \approx M_p^3/m_N^2 \approx 1.8 M_\odot$. This mass
estimate is exactly in the region of maximum masses presently discussed
in the literature with much more refined theoretical models of neutron
star matter. There are excellent modern textbooks on compact stars
available, which discuss the physics of neutron stars and its various
phases in great detail. We refer the interested reader to the books by
Norman Glendenning \cite{Glen_book} and by Fridolin Weber
\cite{Weber_book} for a thorough treatment of this exciting field in the
interplay of physics and astrophysics. Most recently, a new textbook on
compact stars appeared which updates and complements the existing ones
\cite{Haensel_book}. A recent review article on the relation between the
nuclear equation of state and neutron stars can be found in
\cite{Lattimer:2006xb}, a nice popular article on neutron stars in
\cite{Lattimer:2004pg}.

In these lecture notes, we discuss the global properties of neutron
stars, masses and radii, and how those change for different compositions
in their interior. In particular, we address the appearance of new and
exotic phases in dense neutron star matter and how they can modify the
mass-radius diagram for compact stars. In section~\ref{sec:hyperons},
the baryons of the SU(3) octet are treated together in a chiral
SU(3)$\times$SU(3) effective Lagrangian which is motivated from the
approximate symmetries of the underlying theory of strong interactions,
quantum chromodynamics (QCD). When the properties of hyperons are tied
to the known experimental data on hypernuclei, hyperons are present in
neutron star matter above twice normal nuclear matter density affecting
the properties of massive neutron star configurations. In
section~\ref{sec:kaons}, strange mesons, the $K^-$, and its presence in
neutron star matter by forming a Bose-Einstein condensate are addressed.
A relativistic field-theoretic approach is utilised to outline the basic
features of kaon condensation for compact stars. Ultimately, hadrons
have to be described in terms of quark degrees of freedom at high
densities. The phase transition from the chirally broken hadronic phase
to the approximately chirally restored quark matter phase is examined
within perturbative QCD and the MIT bag model to illustrate the main
features of a strong first order phase transition in compact star matter
in section~\ref{sec:quarks}. Characteristics of the mass-radius relation
are highlighted which signal the presence of exotic matter in the core
of compact stars.

\section{Hyperons in Neutron Stars}
\label{sec:hyperons}

QCD with massless quarks is chirally symmetric, which means that all
left-handed and right-handed quarks decouple. This statement is actually
true for all vector interactions. Left- and right-handed quarks are
defined by the relations
\begin{eqnarray}
q_L &=& \frac{1}{2} (1-\gamma_5)q \qquad \sim \qquad (3,0)\\ \nonumber 
q_R &=& \frac{1}{2} (1+\gamma_5)q \qquad \sim \qquad (0,3) \qquad .
\end{eqnarray}
Splitting a spinor $\Psi$ to
left and right-handed components one gets
\begin{equation}
\bar\Psi \gamma_\mu A^\mu \Psi = 
(\bar L + \bar R)\gamma_\mu A^\mu (L + R) = 
\bar L \gamma_\mu A^\mu L  + 
\bar R \gamma_\mu A^\mu R  \quad .
\end{equation}
The mass term for quarks violates chiral symmetry as
\begin{equation}
m \bar \Psi \Psi = m (\bar L + \bar R) (L + R) = m(\bar L R + \bar R L)
\end{equation}
and quarks with different chirality mix with each other. The explicit
breaking of chiral symmetry is small, however, as $m_{u,d}\approx
10$~MeV and also $m_s\approx 100$~MeV is much smaller than the nucleon
mass $m_N \approx 1$~GeV so that chiral symmetry is a useful tool. QCD
has a complex and nontrivial structure, quark and gluon condensates are
present in the vacuum, in particular the light quark condensate $\sigma$,
the strange quark condensate $\zeta$ and the gluon condensate $\chi$. The
nonvanishing vacuum expectation values of these condensates actually
generate most of the masses of the hadrons (except for the
pseudo-Goldstone bosons).

In constructing an effective chiral Lagrangian for the description of
neutron star matter, one has to consider composite quark fields for the
meson and baryon fields. First, consider the spin zero mesons. Assuming
that they are $s$-wave bound states, then the only spinless objects we
can form are
\begin{equation}
\overline{q}_R   q_L  \qquad   \overline{q}_L q_R \qquad .
\end{equation}
The combinations $\overline{q}_L q_L$ and $\overline{q}_R q_R$ vanish,
since the left and right chiral subspaces are orthogonal to each other.
The resulting representation in chiral $SU(3)\times SU(3)$ symmetry is
then (3,3$^{\ast}$) and (3$^{\ast}$,3), respectively. The antiparticles
belong to the conjugate representation. Hence, nonets of pseudoscalar
and scalar particles have to be considered.  For the vector mesons, one
has to construct vector-like quantities out of the quark fields $q_L$
and $q_R$. Again assuming s-wave bound states, the only vectors which
can be formed are
\begin{equation}
 \overline{q}_L \gamma_{\mu} q_L  \qquad \overline{q}_R \gamma_{\mu} q_R
 \qquad . 
\end{equation}
This suggests assigning the vector and axial vector mesons to the
representation ($3 \times 3^{\ast}$,0) $\oplus$ (0,$3 \times 3^{\ast}$)
= (8,1) $\oplus$ (1,8), an octet and a singlet state.

\begin{figure}
\centerline{\includegraphics[width=0.7\textwidth]{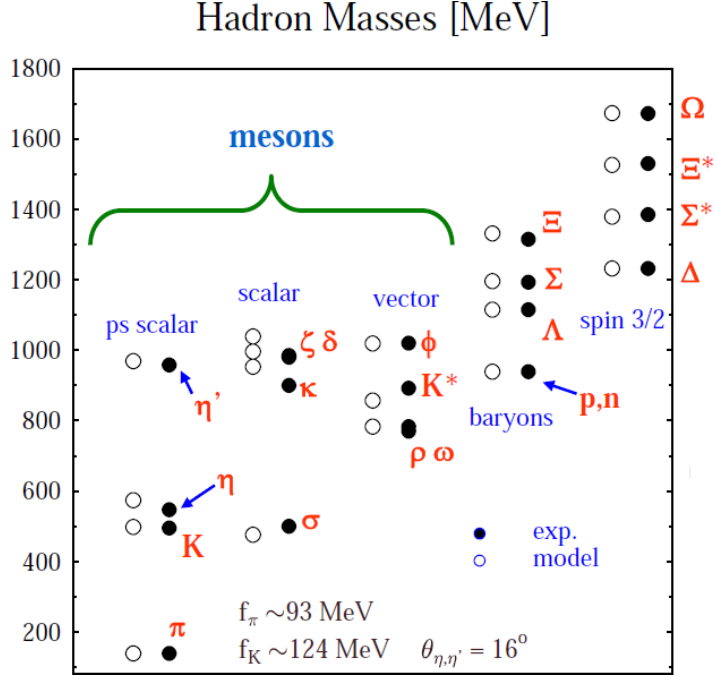}}
\bigskip
\caption{The hadron masses as generated by the vacuum expectation values
  of the chiral SU(3) model in comparison to the experimental data.}
\label{fig:chiral_masses}
\end{figure}

The meson fields can be grouped conveniently into matrices under flavour
SU(3), here for the scalar fields and pseudoscalar fields
\begin{equation}
\sum_{a=0}^8 (\overline{q}_L \lambda^a q_R+\overline{q}_L \lambda^a
\gamma_5 q_R) \equiv \sum_{a=0}^8 (\xi_a \lambda_a +i \pi_a \lambda_a)
= \Sigma+i \Pi = M  
\end{equation}
and correspondingly for the vector and axial vector fields as well as
for the baryon fields. Out of all these fields one constructs a
Lagrangian which obeys the chiral SU(3) symmetry 
\begin{equation}
{\cal L} = {\cal L}_{\rm kin} + {\cal L}_{\rm BM} + {\cal L}_{\rm BV} +
{\cal L}_{\rm vec} + {\cal L}_0 + {\cal L}_{\rm SB} + {\cal L}_{\rm lep} ~,
\end{equation}
with the usual kinetic terms for baryons (${\cal L}_{\rm kin}$), spin-0
fields (${\cal L}_0$), vector mesons (${\cal L}_{\rm vec}$) and leptons,
electrons and muons (${\cal L}_{\rm lep}$). The baryons couple with
Yukawa-type interactions to the spin-0 mesons (${\cal L}_{\rm BM}$) and
to spin-1 mesons (${\cal L}_{\rm BV}$). An explicit chiral symmetry
breaking term is introduced also (${\cal L}_{\rm SB}$). For homogeneous
matter and in the mean-field approximation, derivatives of the boson
fields vanish and the fields with unnatural parity (pseudoscalars and
axial vector fields) vanish. One arrives at the following expressions
for the various terms in the Lagrangian:
\begin{eqnarray}
{\cal L}_{\rm BM}+{\cal L}_{\rm BV} &=& -\sum_{i} \overline{\psi}_i \left[
m^*_i + g_{i \omega}\gamma_0 \omega^0 + g_{i \phi}\gamma_0 \phi^0
+ g_{N\rho} \gamma_0 \tau_3 \rho_0 \right] \psi_{i} ~, \nonumber \\
{\cal L}_{\rm vec} &=& \frac{1}{2} m_{\omega}^2\frac{\chi^2}{\chi_0^2}
\omega^2 + \frac{1}{2}  m_\phi^2\frac{\chi^2}{\chi_0^2} \phi^2
+ \frac{1}{2} \frac{\chi^2}{\chi_0^2} m_{\rho}^{2}\rho^2
+ g_4^4 (\omega^4 + 2 \phi^4 + 6 \omega^2 \rho^2 +\rho^4 ) ~, \nonumber \\
{\cal L}_0 &=& -\frac{1}{2} k_0 \chi^2 (\sigma^2+\zeta^2)
+ k_1 (\sigma^2+\zeta^2)^2 + k_2 ( \frac{ \sigma^4}{2} + \zeta^4)
+ k_3 \chi \sigma^2 \zeta \nonumber \\
& & - k_4 \chi^4 - \frac{1}{4}\chi^4 \ln \frac{ \chi^4 }{ \chi_0^4}
+\frac{\delta}{3}\ln \frac{\sigma^2\zeta}{\sigma_0^2 \zeta_0} ~,\nonumber \\
{\cal L}_{\rm SB} &=& -\left(\frac{\chi}{\chi_0}\right)^2
\left[m_\pi^2 f_\pi \sigma + (\sqrt{2}m_K^2 f_K - \frac{1}{\sqrt{2}}
m_{\pi}^2 f_{\pi})\zeta \right] ,\nonumber \\
{\cal L}_{\rm lep} &=& \sum_{l=e, \mu} \overline{\psi}_l
[i \gamma_\mu\partial^\mu - m_l ]\psi_l ~.
\end{eqnarray}
Nonvanishing vacuum expectation values are generated by spontaneous
breaking of chiral symmetry, so there appears a light quark condensate
and a strange quark condensate. In addition, effects from the
nonvanishing gluon condensate are taken into account by introducing a
scalar $\chi$ field, which is a chiral singlet. These condensates
generate the hadron masses. So hadron masses are not additional input
parameters in this effective field theory and come out automatically by
considering chiral SU(3) symmetry, contrary to e.g.\ the standard
relativistic mean-field theory (there is for example no explicit mass
term for baryons in the Lagrangian!). The expressions for the baryon
masses read for example
\begin{eqnarray}
 m_N^* &=& m_0 -\frac{1}{3}g_{O8}^S(4\alpha_{OS}-1)
(\sqrt{2}\zeta-\sigma) \nonumber \\
m_{\Lambda}^* &=& m_0-\frac{2}{3}g_{O8}^S(\alpha_{OS}-1)
(\sqrt{2}\zeta-\sigma) \nonumber \\
m_{\Sigma}^* &=& m_0+\frac{2}{3}g_{O8}^S(\alpha_{OS}-1)
(\sqrt{2}\zeta-\sigma) \nonumber \\
m_{\Xi}^* &=& m_0+\frac{1}{3}g_{O8}^S(2\alpha_{OS}+1)
(\sqrt{2} \zeta-\sigma) \quad .
\end{eqnarray}
with $m_0=g_{O1}^S(\sqrt{2} \sigma+\zeta)/\sqrt{3}$. The generated hadron mass
spectrum is compared to the experimental data in
Fig.~\ref{fig:chiral_masses}.

For determining the necessary input to the Tolman-Oppenheimer-Volkoff
(TOV) equation, the relation between pressure and energy density has to be
calculated. The thermodynamic grandcanonical potential
$$
\Omega/V = -{\cal L}_{\rm vec} - {\cal L}_0 - {\cal L}_{\rm SB}
-{\cal V}_{\rm vac} - \sum_i \frac{\gamma_i}{(2\pi)^3} \int
d^3k \left[E_i^*(k) - \mu_i^*\right]
- \frac{1}{3} \sum_l \frac{1}{\pi^2} \int
\frac{ dk \: k^4}{\sqrt{k^2 + m_l^2}} .
$$
can be derived by standard methods. The particle energy of the baryons
depends now on the expectation values of the meson fields in the medium
\begin{equation}
E_i^*(k_i) = \sqrt{k_i^2 + m_i^{* 2}}
\end{equation}
so that the baryons acquire an effective mass $m^*$ and an effective
chemical potential
\begin{equation}
\mu_i = E_i^*(k_{F,i}) + g_{i\omega}\omega_0
+ g_{i\phi}\phi_0  + g_{i\rho}I_{3i}\rho_0 ~,
\end{equation}
which is shifted by the vector potentials. All thermodynamic quantities,
as the number density $n$ and the energy density $\epsilon$, can be
extracted from the thermodynamic potential via
\begin{equation}
p = -\frac{\Omega}{V} \quad , \qquad n_i = \frac{\partial p}{\partial
  \mu_i} 
\quad, \qquad \epsilon = -p + \sum_i \mu_i n_i \quad .
\end{equation}
The coupling constants of the scalar mesons to the baryons are fixed by
determining the baryon masses in vacuum. The vector coupling constant
are automatically given by SU(3) symmetry relation:
\begin{equation}
g_{\Lambda \omega} = g_{\Sigma \omega} = 2 g_{\Xi \omega}
= \frac{2}{3} g_{N \omega}=2 g_{O8}^V ; \qquad
g_{\Lambda \phi} = g_{\Sigma  \phi} = \frac{g_{\Xi \phi}}{2} =
\frac{\sqrt{2}}{3} g_{N \omega} ~.
\end{equation}
which are actually the SU(6) symmetry relations known from the quark
model (ideal mixing is assumed which is a very good approximation for
the vector meson nonet).

\begin{figure}
\vspace*{-1cm}
\centerline{\includegraphics[width=1.2\textwidth]{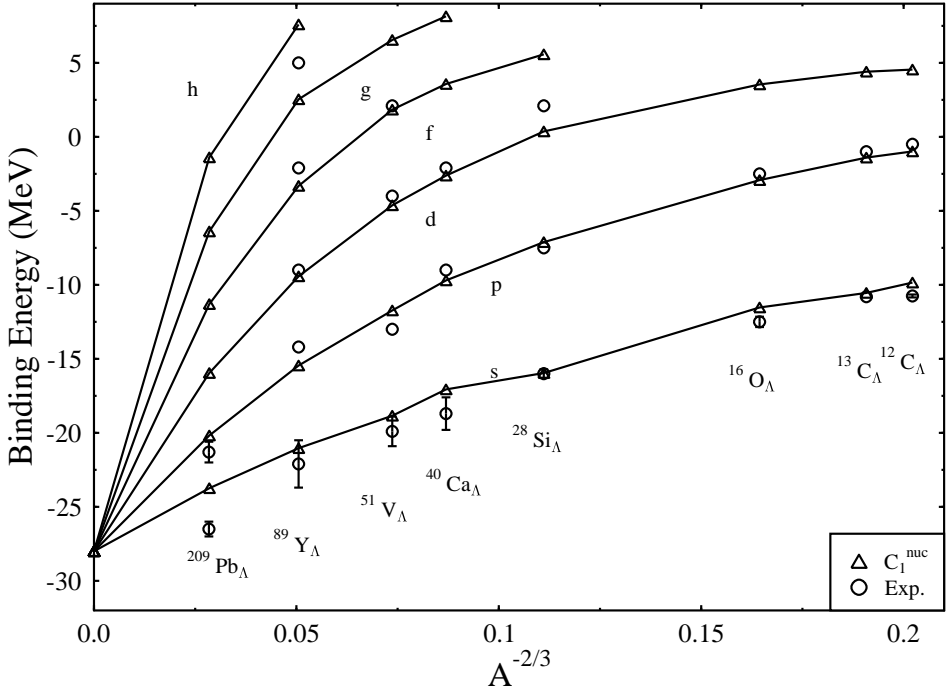}}
\vspace*{-0.5cm}
\caption{The single-particle energy levels of various hypernuclei
  for the SU(3) chiral model compared to the experimental data (taken
  from \cite{Beckmann:2001bu}).}
\label{fig:hypernuclei}
\end{figure}

The chiral effective model gives a good description of nuclear matter as
well as of the properties of nuclei. The nuclear matter properties are a
binding energy of $E_B/A=-16$~MeV at $n_0 = 0.15$ fm$^{-3}$ with an
effective mass of $m^*_N/m_N = 0.61$, a compression modulus of $K=276$
MeV, and an asymmetry term of $a_{\rm sym} = 40.4$ MeV. Hyperons are
automatically included by adopting consistently SU(3) symmetry for the
chiral effective Lagrangian from the beginning. The computed
single-particle energy levels of hypernuclei are depicted in
Fig.~\ref{fig:hypernuclei} and compared with the experimental
hypernuclear data. One sees an overall agreement with the data, just
reflecting the fact that $\Lambda$ hypernuclei are well described by a
potential depth of the $\Lambda$ at saturation density of about
$-30$~MeV. The situation for the other hyperons, $\Sigma$ and $\Xi$, is
far less clear. The $\Sigma$ hypernuclear potential is likely to be
repulsive, while there is experimental evidence that the $\Xi$
hypernuclear potential is attractive, although significantly less in
comparison to the one for $\Lambda$ hyperons. In the chiral effective
model, the hyperon potential are fixed already by the parameters and the
SU(3) symmetry relations.  The hyperon potential for the $\Sigma$ comes
out to be only barely repulsive in the chiral model, which will be
important for the composition of neutron star matter.

\begin{figure}
\vspace*{-1cm}\hspace*{-1cm}
\includegraphics[width=1.0\textwidth]{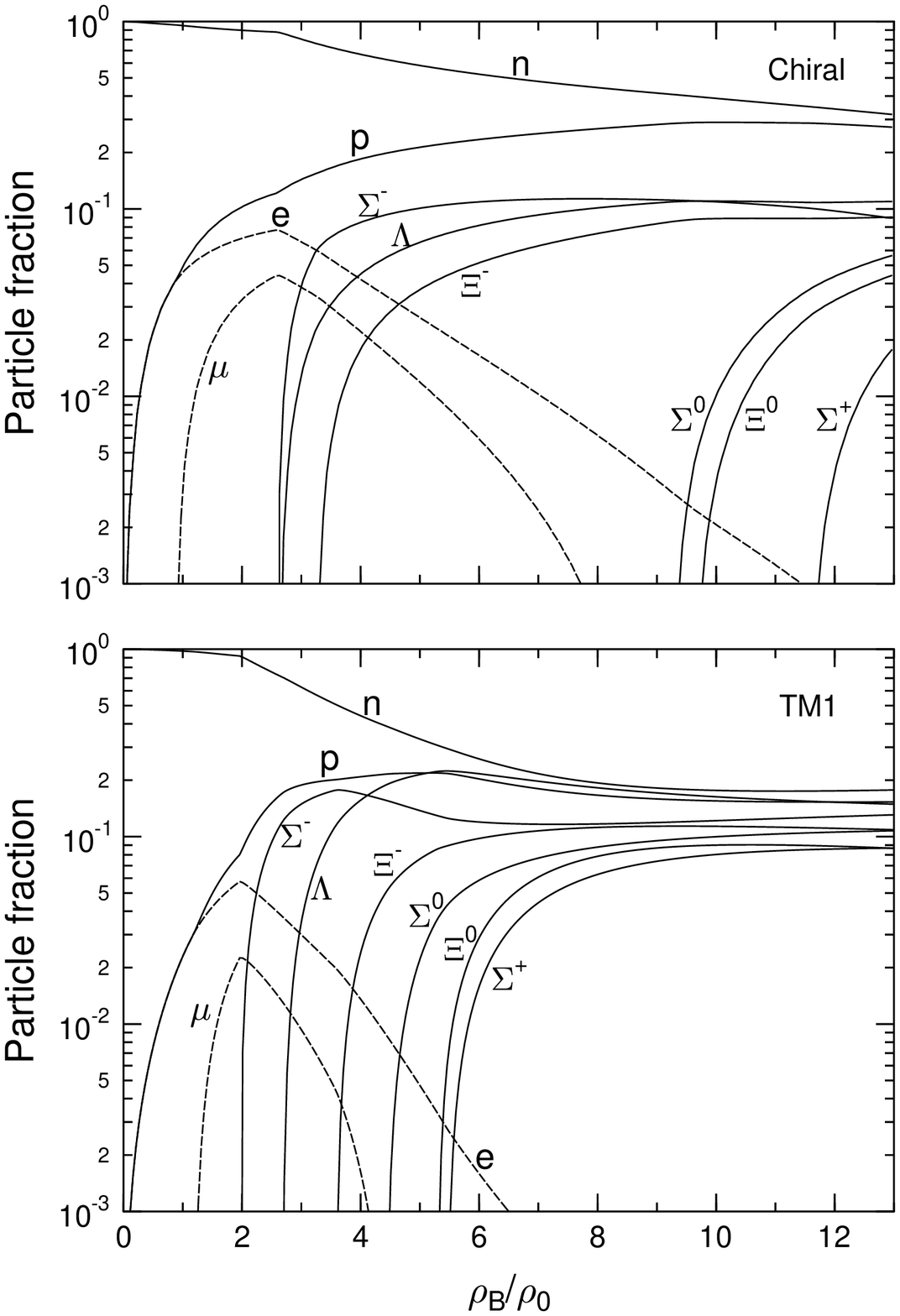}
\vspace*{-3cm}
\caption{The composition of neutron star matter as a function of density
  for the chiral SU(3) model (upper plot) and the relativistic
  mean-field model using the parameter set TM1 (lower plot). Hyperons
  appear around twice normal nuclear matter density (taken from
  \cite{Hanauske:1999ga}).}
\label{fig:comp}
\end{figure}

Hyperons can appear in neutron star matter by virtue of the conditions
of $\beta$-equilibrium \cite{Ambart60,Glendenning:1984jr}. The timescale
for weak interaction rates is of the order of $10^{-10}$ s (decay
timescale for hyperons) to $10^{-8}$ s (decay timescale for kaons) in
vacuum. In matter, similar rates are expected, as for example hyperons
in hypernuclei have similar to just slightly smaller lifetimes compared
to the hyperon lifetime in vacuum.  Neutron stars are known to be quite
old, some of them have characteristic ages close to the age of the
universe of $10^{10}$ years, plenty of time to reach equilibrium for
weak interactions, $\beta$-equilibrium. Hence, $\Lambda$ hyperons can
appear in dense neutron star matter if their effective energy in the
medium reaches the baryochemical potential of neutrons at some baryon
density $n$: $E^*_\Lambda(n)=\mu_\Lambda=\mu_n$. The relations between
the chemical potentials of all particles are determined by the conserved
charges of the particles, which for neutron star matter are just baryon
number and charge (not strangeness, contrary to e.g.\ heavy-ion
reactions). Hence, the chemical potential of all particles can be fixed
by
\begin{equation}
\mu_i = B_i \cdot \mu_B + Q_i \cdot \mu_Q
\end{equation}
where $B_i$ and $Qi$ are the baryon number and the charge of the
particle $i$, respectively. The effective energy for hyperons increases
less rapidly than the one for nucleons, due to the SU(3) symmetry of the
vector coupling constant of the vector potential, which dominates the
high-density behaviour. Therefore, modern calculations of the
composition of neutron star matter predict that hyperons appear around
$2n_0$, where $n_0$ stands for the normal nuclear matter density.
Fig.~\ref{fig:comp} shows the composition as calculated from the chiral
effective model presented above. Neutrons are the main component at low
densities. Protons and electrons appear then with equal amounts, so as
to conserve charge neutrality. Then at about $2.5n_0$ first the
$\Sigma^-$ then the $\Lambda$ hyperons are present in the matter,
reaching fractions of around 10\% at larger densities. The $\Sigma^-$
appears before the $\Lambda$ despite its heavier mass as negatively
charged particles are favoured in neutron star matter, so as to balance
the positive charge of the protons. The asymmetry energy of nucleons
drives the system to isospin-neutral matter, i.e.\ equal amounts of
neutrons and protons. Therefore, also the $\Xi^-$ hyperons appears at
lower densities than the $\Xi^0$, here slightly above $3n_0$ already.
For comparison, the composition for the standard relativistic mean-field
(RMF) model is plotted also in Fig.~\ref{fig:comp}. Here, equal hyperon
potentials have been adopted for all hyperons. The pattern of the onset
of hyperon populations is quite similar in the two models up to $4n_0$.
For larger densities, the other hyperons are present at lower densities
in the RMF model compared to the chiral model. However, the maximum
density reached in neutron star configurations is about $6n_0$, so that
the difference between those two models will be only important for the
most massive neutron star configurations close to the maximum mass.

\begin{figure}
\bigskip
\centerline{\includegraphics[width=0.65\textwidth]{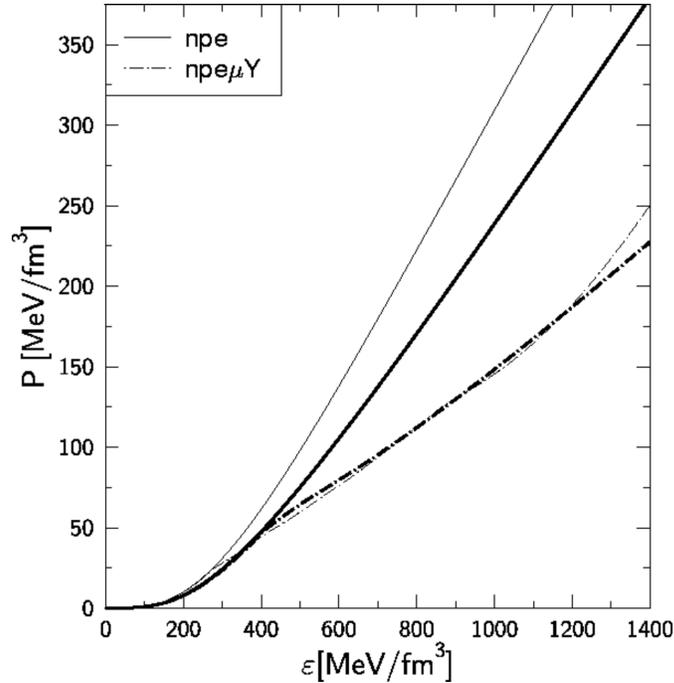}}
\caption{The equation of state of neutron star matter in
  $\beta$-equilibrium with and with out hyperons (taken from
  \cite{Hanauske:1999ga}). Thick lines show the results for the chiral
  model, thin lines the ones for the RMF model using set TM1. Hyperons
  substantially reduce the pressure for a given energy density, i.e.\ 
  they soften the equation of state.}
\label{fig:eos}
\end{figure}

Hyperons have a dramatic effect on the nuclear equation of state (EoS),
the relation between the pressure and the energy density of matter,
which serves as the crucial input to the structure equations for compact
stars, the Tolman-Oppenheimer-Volkoff equations. Fig.~\ref{fig:eos}
shows the equation of state for the chiral model (thick lines) and the
RMF model (thin lines). The upper curves are calculated with nucleons
and leptons only, the lower ones include effects from the presence of
hyperons in dense matter. One sees, that the hyperons substantially
lower the pressure for a given energy density. This effect is stronger
than the difference between the two models used here. In particular, the
high-density EoS with hyperons is nearly the same in both models.

\begin{figure}
\bigskip
\centerline{\includegraphics[width=0.65\textwidth]{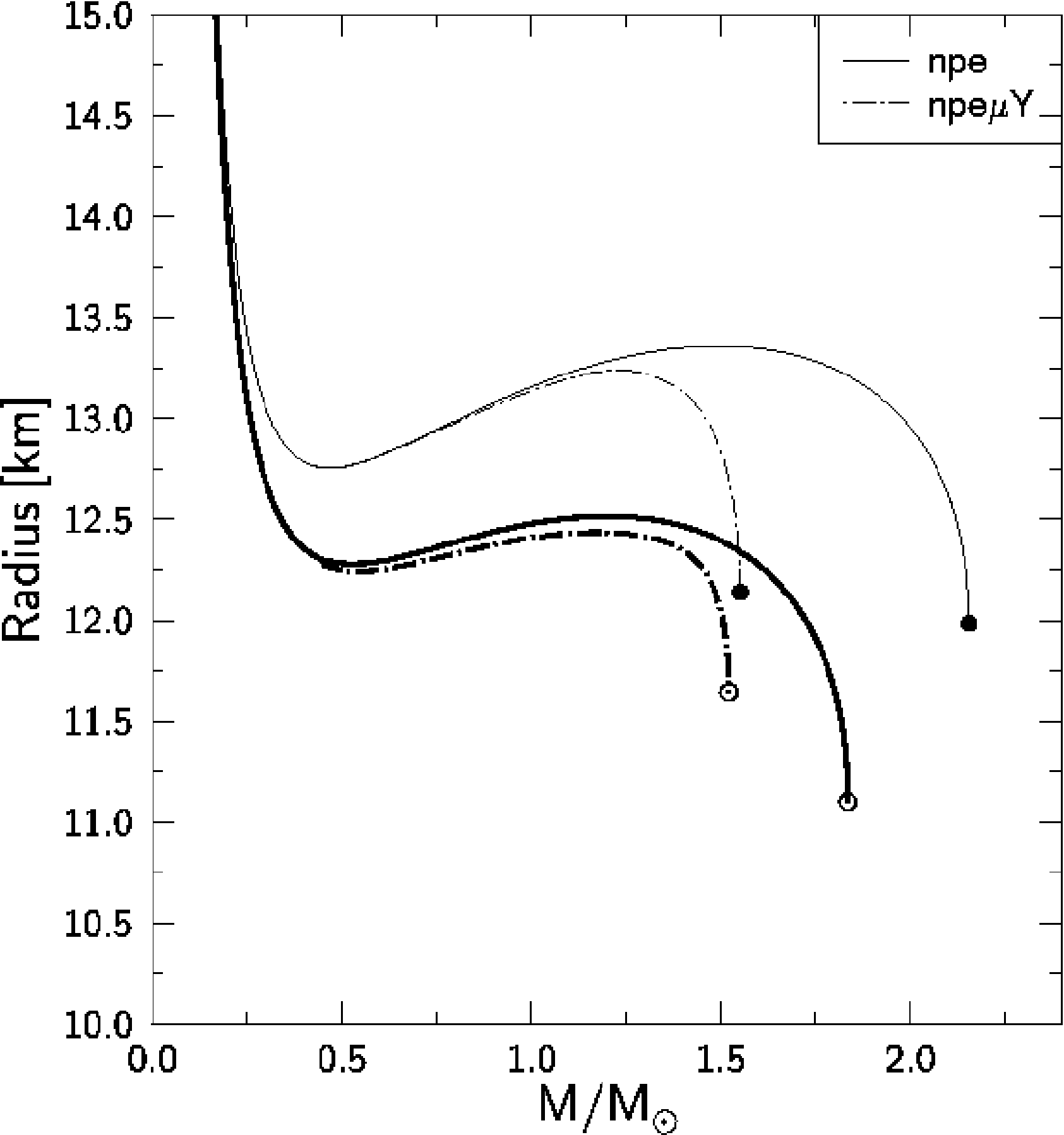}}
\caption{The mass-radius relation for the relativistic mean-field model
  using the parameter set TM1 (thin lines) and for the chiral SU(3)
  model (thick lines) with and without hyperons (taken from
  \cite{Hanauske:1999ga}). Hyperons substantially reduce the maximum
  possible mass.}
\label{fig:m-r}
\end{figure}

The lowering of the pressure has a destablizing effect for neutron
stars, as pressure is needed to counteract the attractive pull of
gravity. Hence, one expects that the maximum possible mass for compact
stars with hyperons is significantly lower than the one for neutron
stars with just nucleons and leptons. Fig.~\ref{fig:m-r} demonstrates
this effect of hyperons on the mass-radius diagram of compact stars. 
Matter with nucleons and leptons only arrive at maximum masses of
$1.84M_\odot$ for the chiral and $2.16M_\odot$ for the RMF model,
respectively. The maximum masses with hyperons included are reduced to
$1.52M_\odot$ for the chiral and $1.55M_\odot$ for the RMF model. Note,
that the radii are different already at lower masses which is due to the
difference in the EoS slightly above normal nuclear saturation
density. The drastic reduction of the maximum mass due to hyperons is
known to be a quite generic feature, see e.g.~\cite{Glendenning:1991es}.

\begin{figure}
\bigskip
\centerline{\includegraphics[width=0.65\textwidth]{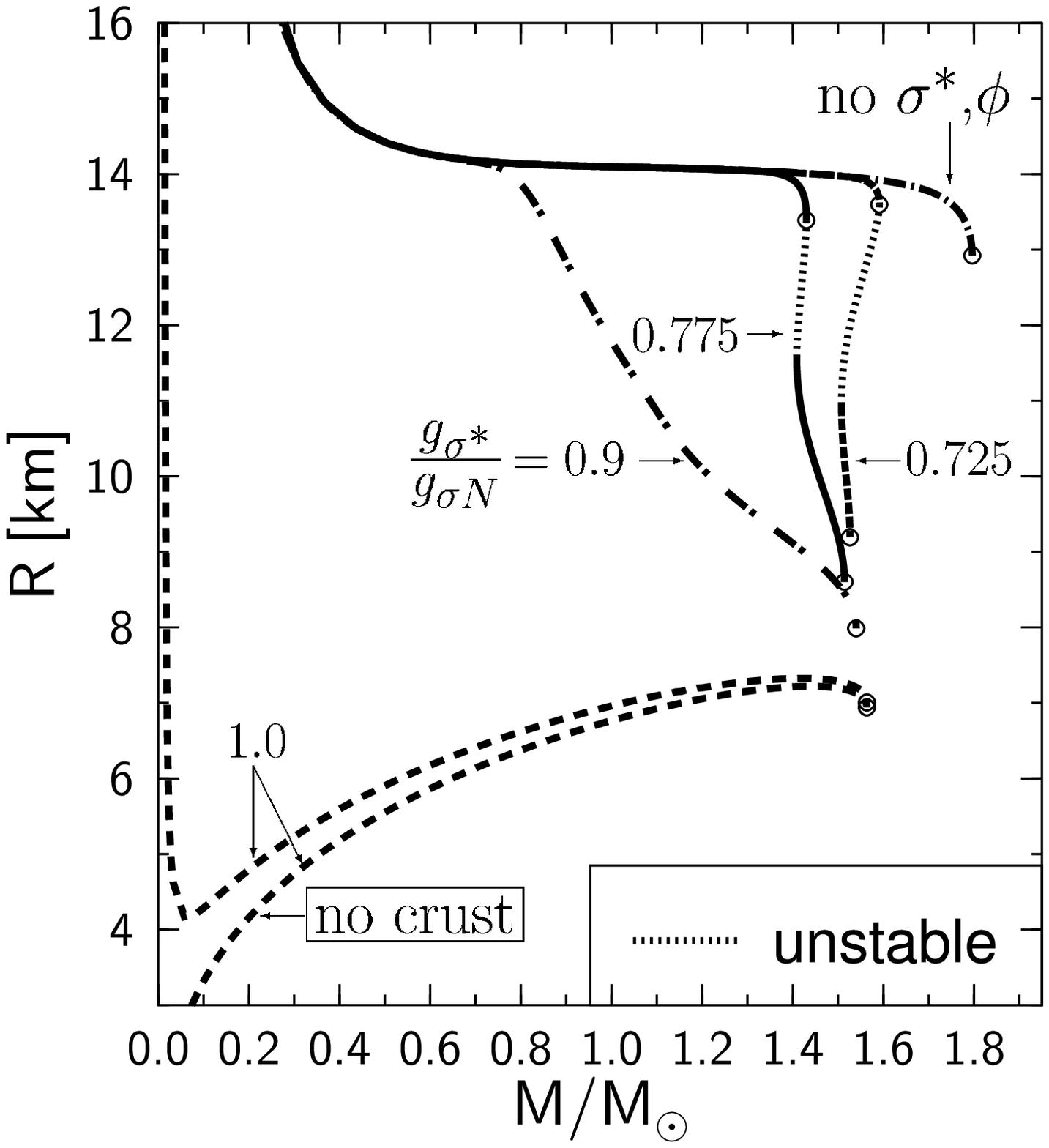}}
\caption{The mass-radius relation for different strength of the
  attractive hyperon-hyperon interactions as given by the coupling
  constant $g_{\sigma*}$ (taken from \cite{SchaffnerBielich:2002ki}). A
  new stable solution appears in the mass-radius relation with similar
  masses but smaller radii. Selfbound hyperon stars are shown by
  long-dashed lines.}
\label{fig:m-r-yy}
\end{figure}

Finally, the transition to hyperon matter is studied by tuning up the
hyperon-hyperon interaction, which is scarcely known from the few double
$\Lambda$ hypernuclear events available. With increasing hyperon-hyperon
attraction, the transition to hyperons becomes a first order phase
transition. The mixed phase, with slowly rising pressure as a function
of energy density, will make the compact star less stable. The onset of
the pure hyperonic phase with a steeply rising pressure can lead to
another stable sequence of compact stars. For a sufficiently huge
attraction between hyperons, hyperonic matter becomes more stable than
pure nucleonic neutron star matter and the mass-radius relation changes
to the one for selfbound stars, i.e.\ the mass increases from the origin
with the radius as $R^3$, so that the average density in the star is
nearly constant. Fig.~\ref{fig:m-r-yy} depicts the mass-radius
diagram for an increased hyperon-hyperon attraction. For moderate
attraction, the mass-radius relation has two distinct branches with
correspondingly two maximum masses. The new branch at smaller radii is
another stable solution to the TOV equations, constituting the so called
third family of compact stars \cite{Glendenning:1998ag,Schertler:2000xq}.
For even larger hyperon attraction, the mass-radius relation changes to
the one for selfbound stars: the mass-radius relation starts at the
origin. Selfbound stars can still possess an outer nuclear crust, which
is determined by the low-density nuclear EoS below neutron-drip density.
In that case, the mass-radius relation for the low-mass configurations
interpolates between the selfbound star configuration and the normal
low-mass neutron star sequence.

\section{Strange Bosons in Hadron Stars}
\label{sec:kaons}

Besides hyperons, other hadronic particles can be present at high
density in the core of neutron stars. In the following, we discuss how
bosons and the phenomenon of Bose-Einstein condensation can be described
in a simple relativistic field-theoretical approach, here for the case
of kaon condensation in neutron star matter
\cite{Kaplan:1986yq,Brown92}. We follow the field-theoretical model of
Ref.~\cite{Glendenning:1998zx,Glendenning:1997ak}.

In dense matter, kaons, and in particular the negatively charged $K^-$,
can appear in neutron star matter. For that to happen, it must be
energetically favoured to replace electrons by $K^-$ meson which
translates to the condition that the effective energy of the $K^-$ in
the dense medium must be equal to the chemical potential of electrons, 
$E_K^*(n)=\mu_e$. The in-medium shift of the mass and the energy of the
kaons can be modelled by Yukawa coupling terms to scalar and vector
fields generated by the nuclear matter.
Here, for simplicity, we adopt the standard relativistic mean-field
model for the nuclear part and just add those coupling terms for the
kaon field to the Lagrangian:
\begin{eqnarray}
{\cal L} &=& 
  \sum_B \bar \Psi_B \left(i \gamma_\mu \partial^\mu - m_B + g_{\sigma B}
  \sigma - g_{\omega B}
  \gamma_\mu V_\mu - g_{\rho B} \vec{\tau}_B \vec{R}_\mu \right) \Psi_B
+ \frac{1}{2} \partial_\mu \sigma \partial^\mu \sigma - \frac{1}{2}
m_\sigma^2 \sigma^2 \cr
&&  - U(\sigma)
- \frac{1}{4} V_{\mu\nu} V^{\mu\nu} + \frac{1}{2}m_\omega^2 V_\mu V^\mu
+ U(V) 
- \frac{1}{4} \vec{R}_{\mu\nu} \vec{R}^{\mu\nu}
+ \frac{1}{2}m_\rho^2 \vec{R}_\mu \vec{R}^\mu
\end{eqnarray}
where the selfinteraction terms between the meson fields read
\begin{equation}
U(\sigma) = \frac{1}{3} bm (g_\sigma \sigma)^3
          + \frac{1}{4} c (g_\sigma \sigma)^4\,, \qquad
U(V) = \frac{d}{4} (V_\mu V^\mu)^2\,.
\end{equation}
If the critical condition for the onset of kaon condensation is fulfilled:
\begin{equation}
\omega_K = \mu_{K^-} = \mu_e
\end{equation}
then processes like
\begin{equation}
e^- \to K^- + \nu_e \qquad n \to p + K^-
\end{equation}
produce kaons in the dense medium. The Lagrangian for the kaon field can
be cast in the form
\begin{equation}
{\cal L}_K = {\cal D}_\mu^* K^* {\cal D}^\mu K - {m^*_K}^2 K^*K
\end{equation}
where the vector fields are coupled minimally
\begin{equation}
{\cal D}_\mu = \partial_\mu + i g_{\omega K} V_\mu +
i g_{\rho K} \vec{\tau}_K \vec{R}_\mu
\end{equation}
and the effective mass of the kaon is defined as a linear shift of the mass
term by the scalar field
\begin{equation}
m^*_K = m_K - g_{\sigma K} \sigma
\quad .
\end{equation}
The in-medium effective energy of the kaon can be read off from the
energy-momentum relation by using a plane wave ansatz for the kaon
field:
\begin{equation}
\omega_K = m_K - g_{\sigma K}\sigma -
g_{\omega K} V_0 - g_{\rho K} R_{0,0}
\quad .
\end{equation}
The total energy density and pressure can be calculated with standard
techniques from the energy-momentum tensor (assuming an ideal fluid), so
that
\begin{eqnarray}
\epsilon &=& \epsilon_N  + \epsilon_K  + \epsilon_{e,\mu} \\
p &=& p_N + p_{e,\mu}
\quad .
\end{eqnarray}
The explicit expressions for the energy density originating from
kaon condensation reads simply
\begin{equation}
\epsilon_K = m^*_K n_K
\quad .
\end{equation}
where $n_K$ stands for the number density of kaons. Note, that there
is no direct contribution from the kaon condensate to the pressure, as
it vanishes for a Bose-Einstein condensate. Finally, one has to fix the
coupling constants for the kaons. We adopt again symmetry arguments to
relate the vector coupling constants for the kaons to the ones of the nucleon
\begin{equation}
g_{\omega K} = \frac{1}{3} g_{\omega N} \quad \mbox{ and } \quad
g_{\rho K} = g_{\rho N}
\quad.
\label{eq:iso_coupl}
\end{equation}
Here, the coupling to the vector mesons are assumed to be universal,
which is the case when gauging the Lagrangian, so that the coupling
constants are just related by isospin symmetry. An extension to SU(3) is
straightforward, but not necessary here, as we discuss first neutron
star matter with nucleons, leptons and kaons only and ignore the effects
from hyperons. The scalar coupling constant can be related to the
relativistic kaon in-medium potential via
\begin{equation}
U_K (\rho_0) =  -g_{\sigma K} \sigma( \rho_0) - g_{\omega K} V_0 (\rho_0)
\end{equation}
which is a more appropriate (and pragmatic) control parameter. The kaon
potential is not well known, but likely to be considerably attractive at
high densities. Coupled channel calculations for kaons in dense matter
arrive at values of up to $-120$~MeV at normal nuclear matter density
\cite{Koch94,Waas96b,Tolos01} although much shallower potentials are
found in selfconsistent treatments, see e.g.~\cite{Lutz98,Tolos:2006ny}.

\begin{figure}
\hspace*{2.5cm}\includegraphics[width=0.9\textwidth]{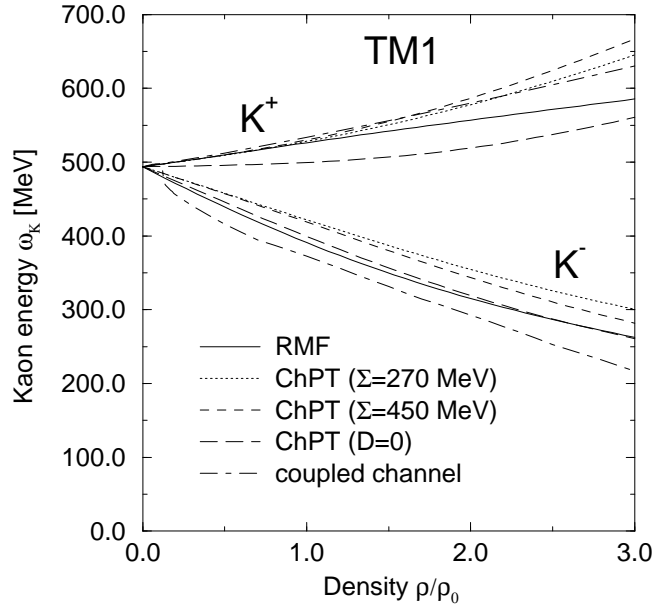}
\caption{The kaon and antikaon energy in dense nuclear matter (taken
  from \cite{Schaffner:1996kv}). The kaon energy is shifted up, while
  the antikaon energy is greatly reduced as a function of density due to
  an overall strongly attractive antikaon-nucleon potential. RMF:
  relativistic mean-field model, CHPT: chiral perturbation theory with
  different values of the $\Sigma$ term and for a vanishing range term,
  $D=0$ (see \cite{Schaffner:1996kv} for details).}
\label{fig:kaon_energy}
\end{figure}

Fig.~\ref{fig:kaon_energy} shows the in-medium energy of kaons and
antikaons as a function of baryon density for different approaches. The
model outlined here is denoted by the label RMF. The prediction from the
RMF approach are in good agreement with the one from chiral perturbation
theory (ChPT), where several cases are shown but not discussed here
in more details. The kaon energy is shifted up in the nuclear medium due
to the repulsive vector potential. The antikaon on the other hand
experiences an attractive vector potential, so that the effective energy
decreases dramatically as a function of density. For comparison, a
coupled channel calculation for the $K^-$ is also plotted, with rather
similar results to the ones of the RMF model.

\begin{figure}
\bigskip
\centerline{
\includegraphics[width=0.7\textwidth]{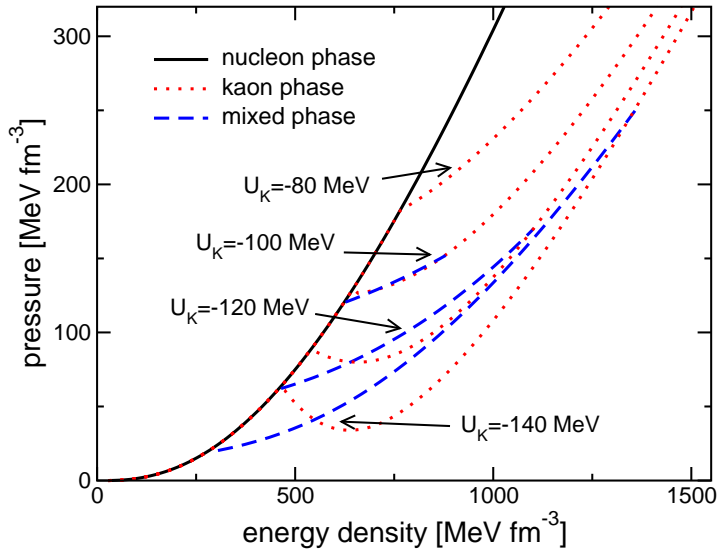}}
\bigskip
\caption{The equation of state for nucleon star matter with kaon
  condensation (from \cite{Glendenning:1997ak}). For large attractive
  kaon potentials, the transition to kaon condensation is of first
  order. The pressure decreases as function of energy density at the
  onset of kaon condensation (red dotted lines) indicating an
  instability. A Gibbs construction has to be used for the description
  of the mixed phase (blue dashed lines).}
\label{fig:kaon_eos}
\end{figure}

The appearance of kaon condensation in neutron star matter is
accompanied by a strong first order phase transition. The Gibbs criteria
for handling a phase transition with two conserved charges, baryon
number and charge, states that the pressure in the two phases has to
be equal for the same chemical potentials
\begin{equation}
p^{\rm I} = p^{\rm II} \,, \quad \mu_B^{\rm I} = \mu_B^{\rm II}\,, \quad
\mu_e^{\rm I} = \mu_e^{\rm II}
\label{eq:gibbscond}
\end{equation}
The equation of state for neutron star matter is depicted in
Fig.~\ref{fig:kaon_eos}. The solid line stands for the purely nucleonic
EoS, the dotted lines for the pure kaon condensed phase with nucleons
and kaons for different values of the kaon potential. The pure kaon
condensed phase is seen to be unstable, there is a region where the
pressure decreases with increasing energy density. The Gibbs
construction for the mixed phase of pure nucleon matter and pure kaon
condensed matter (dashed lines) interpolates between the two phases in a
thermodynamical consistent way, so that the pressure is continuously
rising as a function of energy density.

\begin{figure}
\centerline{
\includegraphics[width=0.8\textwidth]{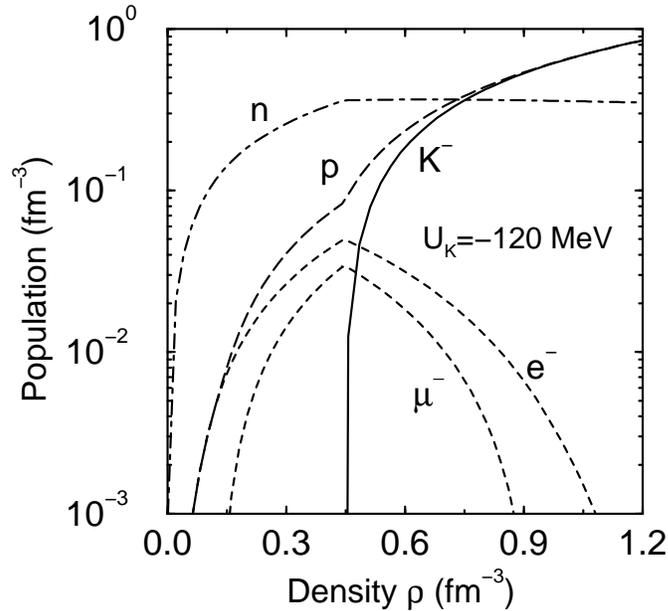}}
\vspace*{-0.5cm}
\caption{The population of nuclear star matter with kaon condensation
  for a kaon optical of $U=-120$ MeV at $n=n_0$ (taken from
  \cite{Glendenning:1997ak}). Kaons appear around $3n_0$ and replace the
  electrons.  The proton fraction balances the negative charge of the
  kaons and can be even larger than the neutron fraction for high
  densities.}
\label{fig:kaon_popul}
\end{figure}

The composition of neutron star matter with kaon condensation is shown
in Fig.~\ref{fig:kaon_popul} for a kaon potential of $U_K=-120$ MeV at
$n_0$.  The $K^-$ set in at $3n_0$ and reach a very high fraction
already for slightly larger densities. As soon as a the $K^-$ are
present, the electron and muon fraction decreases accordingly. At $5n_0$
there are equal fractions of neutrons, protons and $K^-$, so neutron
star matter would be more aptly called nucleon star matter. At even
larger densities, the $K^-$ dominate the population together with the
protons, which ensures overall charge neutrality.  The $K^-$ in
combination with protons are favoured in comparison to the neutrons as
they are deeply bound in dense matter.

\begin{figure}
\vspace*{-1cm}
\centerline{
\includegraphics[width=0.7\textwidth]{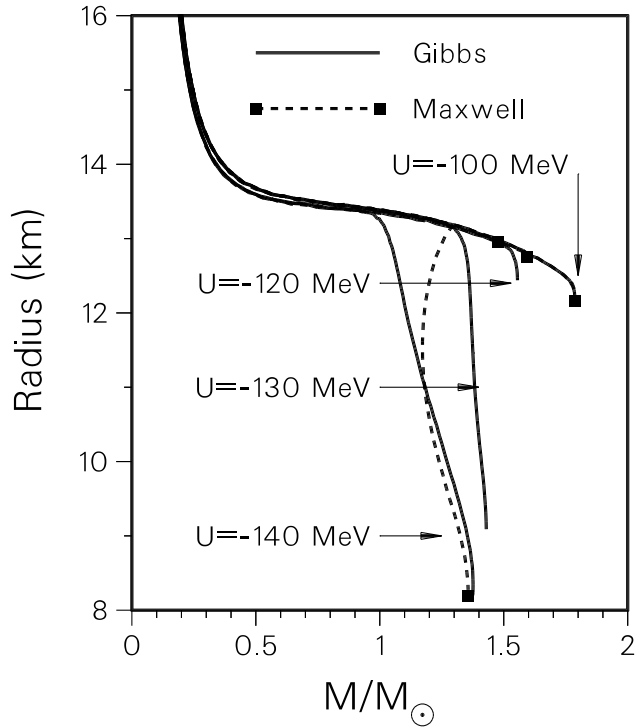}}
\bigskip
\caption{The mass-radius relation for kaon-condensed nucleon stars
  (taken from \cite{Glendenning:1997ak}). The maximum mass is reduced
  and extremely small radii of about 8 km are possible when kaon
  condensation is present.}
\label{fig:mr_kaon}
\end{figure}

Fig.~\ref{fig:mr_kaon} depicts the corresponding mass-radius relation by
using the equations of state as seen in Fig.~\ref{fig:kaon_eos}. The
lower the kaon optical potential, the lower is the critical density for
the onset of kaon condensation. Hence, the mass-radius relation will
deviate from the one of the canonical neutron star at lower masses for
larger values of the kaon potential.  As kaons are in a Bose-Einstein
condensate and do not give a direct contribution to the pressure, less
mass can be supported against the gravitational pull. For low values of
the kaon potential, the mass-radius curve simply gets unstable as soon
as kaons are part of the composition in the core of the compact star.
For the case $U_K=-130$~MeV, the mass slowly changes as a function of
radius and very small radii well below 10 km are reached. High
compression is needed to stabilise a kaon condensate compact star. For
even more attraction, the case $U_K=-140$ MeV, there appears a
significant difference in the mass-radius curves between the
(thermodynamically inconsistent) Maxwell construction and the Gibbs
construction. A strong first order phase transition is present in the
core of the compact star from neutron star matter to kaon condensed star
matter. Interestingly, the deviations in the mass-radius curve due to
the different descriptions of the mixed phase remains to be just in some
intermediate region of the mass-radius curve. For the last sequence of
stable compact star configuration up to the maximum mass, the two
descriptions give about similar results for the mass-radius curve. The
unstable region in the case of the Maxwell construction, indicated by a
change of the slope of the mass-radius diagram, is absent in the case of
the Gibbs construction. Note, that for the Gibbs construction, the kaon
condensed phase appears at much lower values of the mass of the compact
star compared to the Maxwell construction. We will discuss phase
transitions in more detail with regard to the onset of the quark matter
phase at high densities, to which we turn now.

\section{Strange Quark Matter in Compact Stars}
\label{sec:quarks}

In this section, we discuss the properties of compact stars with quark
matter. First, pure quark stars are addressed and the mass-radius
diagram for so-called selfbound stars. Then, the quark matter equation
of state is matched to the one of the low-density hadronic equation of
state. Compact stars with both types of matter, hadronic and quark
matter, are dubbed hybrid stars.

\subsection{General properties of quark stars}

QCD predicts that quarks at large energy scales are asymptotically
free. For high energy-density matter one considers a gas of free quarks
with corrections from one-gluon exchange. The pressure at zero
temperature and finite quark-chemical potential $\mu$ reads for massless
quarks
\begin{equation}
p (\mu)= \frac{N_f \mu^4}{4\pi^2}
\left\{1-2 \left(\frac{\alpha_s}{\pi}\right) 
- \left[G+N_f\ln{\frac{\alpha_s}{\pi}}
+ \left(11-\frac{2}{3} N_f \right) \ln{\frac{\bar\Lambda}{\mu}} \right]
\left(\frac{\alpha_s}{\pi}\right)^2 \right\}
\end{equation}
where $N_f$ stands for the number of flavours (here $N_f=3$). The
constant $G$ is scheme dependent and in the $\overline{\rm MS}$ scheme
given by $G=G_0-0.536N_f+ N_f\ln{N_f}$, $G_0=10.374 \pm 0.13$
\cite{Fraga:2001id}. The renormalisation scale $\Lambda$ will be fixed
to be proportional to the quark-chemical potential. The expression for
the pressure of massive quarks can be found in \cite{Fraga:2004gz}. 

In compact star matter, weak reactions are in equilibrium so that
\begin{eqnarray}
d & \longrightarrow & u+e^- +\bar{\nu}_{e^-}, \\
s & \longrightarrow & u+e^- +\bar{\nu}_{e^-}, \nonumber \\
s + u & \longleftrightarrow & d + u, \nonumber
\end{eqnarray}
and the chemical potentials for down and strange quarks must be equal,
while the one for up quarks is just shifted by the electrochemical
potential:  
\begin{equation}
\mu_s  = \mu_d =  \mu_u+\mu_e.
\end{equation}
Note, that this amounts to baryon and charge number conservation, the
chemical potentials of the quarks can be also expressed in terms of the
baryochemical potential and the electrochemical, as done in the previous
sections. This fact will be beneficial for our discussion of the
matching of the two phases in the next subsection, for example the
connection $\mu_n \equiv \mu_u + 2 \mu_d$ can then be easily made.
From the pressure one can compute the number density $n$ and energy
density $\epsilon$ via the standard
thermodynamic relations for each quark separately
\begin{equation}
n_f(\mu_f) = \frac{d p_f(\mu_f)}{d \mu_f} \quad \mbox{and} \quad
\epsilon_f(\mu_f)= - p_f(\mu_f) + \mu_f \rho_f(\mu_f).
\end{equation}
In addition, pure quark star matter should be charge neutral so that
for the total charge density
\begin{equation}
n_c=\frac{2}{3} n_u -\frac{1}{3} n_d
-\frac{1}{3} n_s -n_e=0
\end{equation}
where $n_i$ stands for th number density of the species $i$.
Finally, the total energy density and pressure of the quark matter is
\begin{eqnarray} 
\epsilon_Q & = & \epsilon_u+\epsilon_d+\epsilon_s+\epsilon_e, \\
p_Q & = & p_u +p_d +p_s +p_e \quad. 
\end{eqnarray}
For comparison, one arrives at the MIT bag model by setting the coupling
constant to zero and shifting the energy density and pressure of the
quark matter by the bag constant $B$ which describes the
non-perturbative aspects of the QCD vacuum:
\begin{equation} 
\epsilon_{\rm bag}  =  \epsilon_{\rm free} + B \qquad
p_{\rm bag}  =  p_{\rm free} - B  
\end{equation}
which can be combined to 
\begin{equation}
p_{\rm bag} = \frac{1}{3}\epsilon_{\rm bag} - \frac{4}{3}B
\end{equation}
as $p_{\rm free} = \epsilon_{\rm free}/3$. 

In the following we consider quark matter of massless up, down, and
strange quarks.  The current strange quark mass of about 100 MeV is
smaller than the scale of the quark-chemical potential of 300 MeV and
more. Corrections are of the order of $(m_s/\mu)^2$ and can be safely
ignored for our purposes.  For the case of three-flavour massless quarks
quark matter consists of equal number densities of up, down and strange
quarks and is charge neutral by itself. Hence, there are no electrons in
such idealised quark matter (in fact the corrections from the finite
strange quark mass are small). The chemical potentials of all three
quark species are the same and just given by the quark-chemical
potential. The quark number density and the energy density is then 
determined by 
\begin{equation}
n = \frac{dp}{d\mu} \qquad 
\epsilon = -p + \mu \cdot n 
\end{equation}
which fixes the quark matter equation of state in parametric form.  The
pressure actually vanishes for some critical chemical potential. Hence,
there is a energy density jump from the vacuum to the energy density of
quark matter. Pure quark matter is then stable at this characteristic
energy density. This feature of the quark matter equation of state can
be easily seen for the MIT bag model but is also present in the
interacting model. The characteristic energy density is then fixed by
the bag constant, as $\epsilon_{\rm bag} = 4B$ for vanishing pressure.

\begin{figure}
\centerline{
\includegraphics[width=0.9\textwidth]{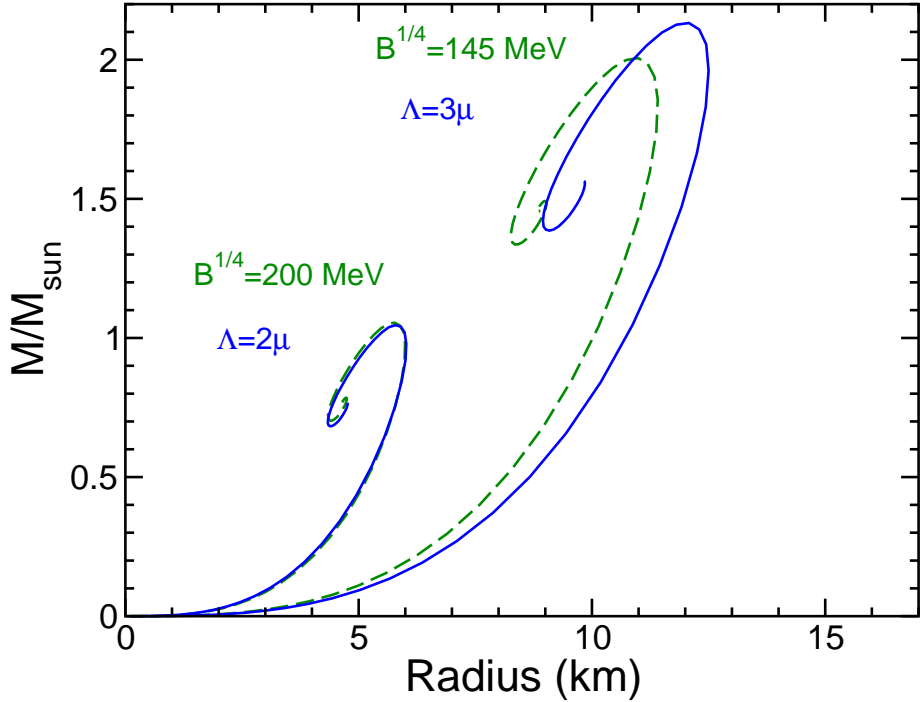}}
\bigskip
\caption{The mass-radius relation of pure quark stars in two different
  approaches: for the MIT bag model (green dashed curves) and within
  perturbative QCD calculation (blue solid lines). The general form of
  the mass-radius relation is remarkably similar.}
\label{fig:mr_quark}
\end{figure}

Fig.~\ref{fig:mr_quark} depicts the mass-radius relation for compact
stars consisting of pure quark matter. The solid lines show the
interacting case for the perturbative QCD equation of state, the dashed
line the one for the MIT bag model for so-called strange stars
\cite{Haensel:1986qb,Alcock:1986hz}. As the pressure vanishes at some
finite value of the energy density, the quark stars are stabilised and
have about similar densities. Hence, the mass increases with the radius
of the quark star as $R^3$ starting at the origin. There is a maximum
mass, as at some critical number of quarks the gravitational pull can
not be counteracted by the Fermi pressure. The curves can be scaled into
each other. For the MIT bag model, maximum mass and the corresponding
radius scale as $B^{-1/2}$, smaller values of the MIT bag constant give
larger maximum masses and larger radii. The result for the perturbative
QCD model are strikingly close to the simple MIT bag model. Here larger
choices for the renormalisation scale result in larger maximum masses
and radii. Note, that pure quark stars can only exist if strange quark
matter is more stable than ordinary nuclear matter, so that any hadronic
mantle is transformed to strange quark matter, which is the so-called
Bodmer-Witten hypothesis \cite{Bodmer:1971we,Witten:1984rs}.

\subsection{Hybrid stars: compact stars with quark and hadron matter}

It is more likely that strange quark matter is not absolutely stable, so
that the quark core is surrounded by hadronic matter. The compact star
is then dubbed a hybrid star. The transition from one phase to the other
in compact star matter has to be modelled via the Gibbs condition for
phase transitions \cite{Glendenning:1992vb} which states that in phase
equilibrium the pressure of the two phases is the same for the same
chemical potentials of the two phases
\begin{equation}
p_H(\mu_B, \mu_e) = p_Q(\mu_B, \mu_e).
\end{equation}
Here, the index $Q$ denotes the quantity in the quark phase, $H$ the one
in the hadronic phase.
There are two conserved charges for compact star matter, the baryon
number and charge, so that there are two independent chemical
potentials. A Maxwell construction could only ensure that one chemical
potential is continuous throughout the phase transition while the other
would jump discontinuously. The volume fraction of the quark phase
\begin{equation}
\chi = \frac{V_Q}{V_Q+V_H}
\end{equation}
can be used to calculate the total energy density in the mixed phase
\begin{equation}
\epsilon_{mix} = \chi \, \epsilon_Q + (1-\chi) \, \epsilon_H.
\end{equation}
The charge neutrality is now guaranteed globally, i.e.\ the Gibbs
construction allows for highly charged phases whose charges balance each
other.

\begin{figure}
\centerline{
\includegraphics[width=0.75\textwidth]{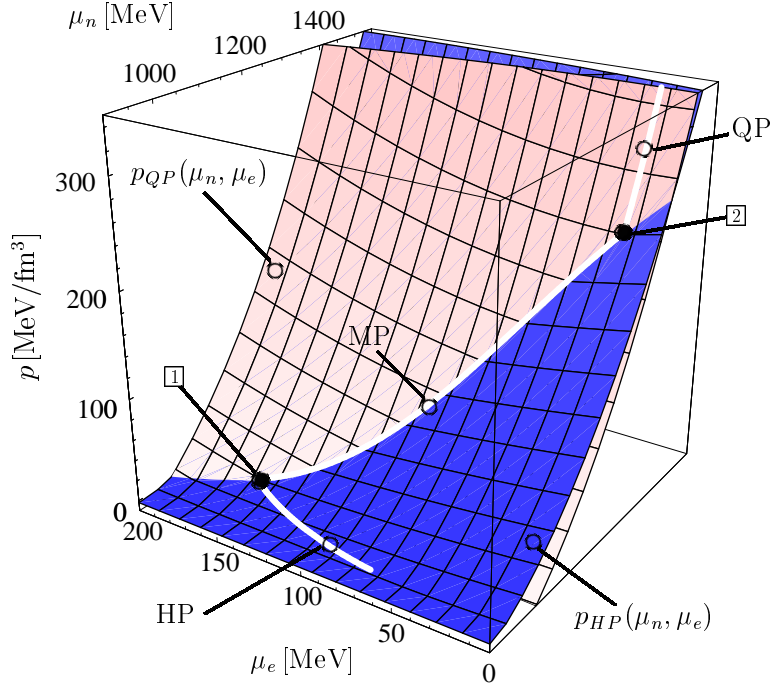}}
\caption{The Gibbs phase construction for two chemical potentials by
  looking at the pressure in 3D. The surface of the hadronic pressure
  cuts the one of the quark matter pressure along the phase coexistence
  line where both pressures are equal. The overall charge neutrality
  condition fixes a line in the hadronic pressure area and also in the
  quark one. Charge neutrality is ensured along the mixed phase line by
  adjusting the volume fractions of hadron and quark matter (taken from
  \cite{Schertler:2000xq}).}
\label{fig:gibbs}
\end{figure}

The Gibbs construction can be visualised by the intersection of the two
planes in a diagram where the pressure is plotted as a function of the
two chemical potentials, see Fig.~\ref{fig:gibbs}. In the plane of the
pressures there is one line which indicates locally charge neutral matter
of a single component, quark or hadron matter. The line of the
intersection of the two pressure planes defines the mixed phase, the
pressure is equal in the two phases. One realizes that the line of
intersection is off the line of charge neutral matter for the quark or
the hadron phase alone. The volume fraction is now chosen in such a
manner that the total mixed phase has a vanishing charge density $\rho$
and is charge neutral globally:
\begin{equation}
\rho_{mix} = \chi \, \rho_Q + (1-\chi) \, \rho_H = 0 .
\end{equation}
At the beginning of the mixed phase, highly negative charged quark matter
bubbles appear while hadron matter is slightly positively charged. At
the end of the mixed phase, quark matter is slightly negative charged
while there are small bubbles of highly positive charged hadron
matter. This mismatch of charges is energetically favoured, hadron
matter is more stable for about equal numbers of protons and neutrons
due to the asymmetry energy, so that a positive charge is advantageous.

\begin{figure}
\centerline{
\includegraphics[width=\textwidth]{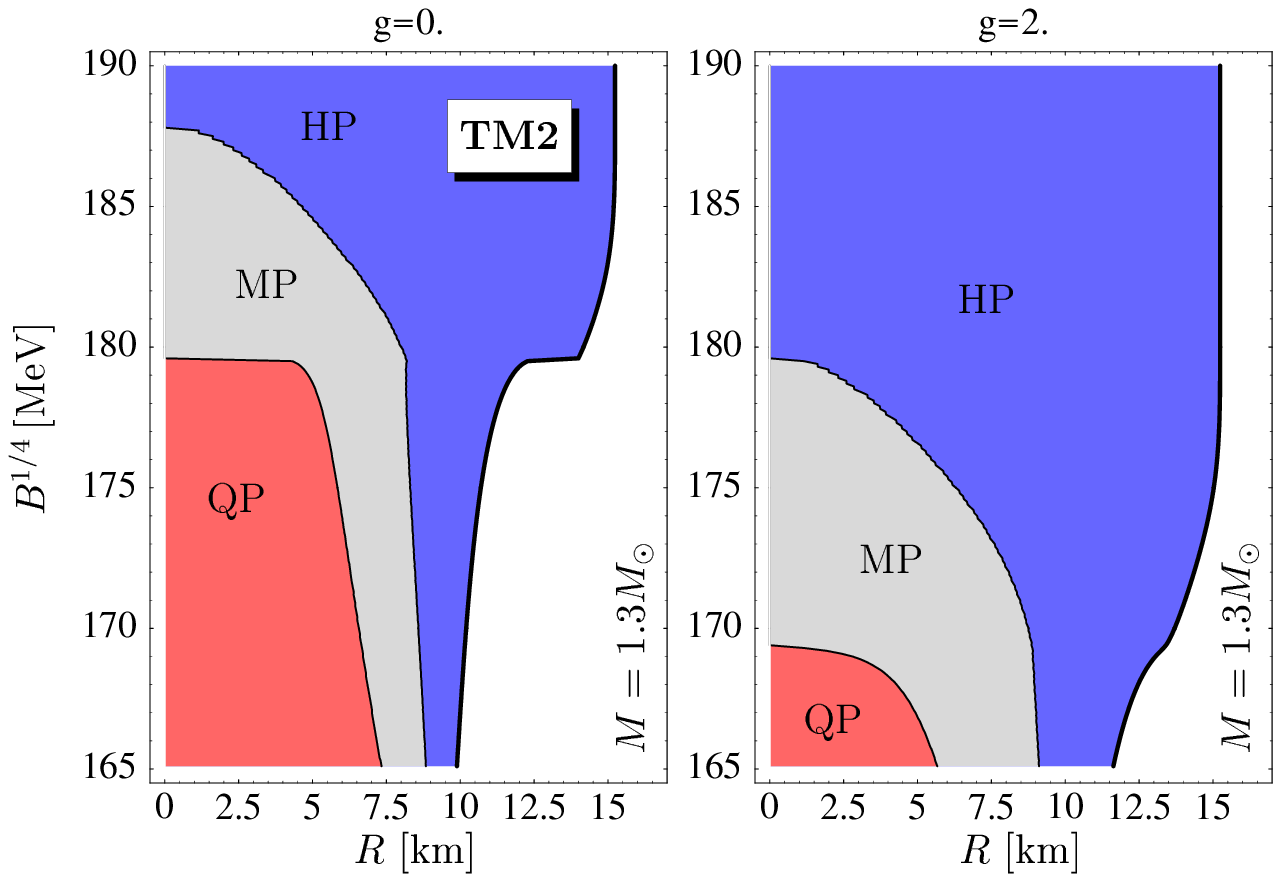}}
\caption{The composition of hybrid stars for different values of the MIT
  bag constant $B$ (taken from \cite{Schertler:2000xq}). For large
  values of $B$, the compact star is purely hadronic. A pure quark core
  is present for small values of $B$. Note, that for even lower values
  of $B$, the pure quark phase will extend up to the surface of the
  compact star and a pure quark star (a so called strange star) is
  formed.}
\label{fig:hybridstar}
\end{figure}

The corresponding composition for hybrid stars is depicted in
Fig.~\ref{fig:hybridstar} as a function of the radius for different
choices of the MIT bag constant. For low values of the bag constant, the
core consists of pure quark matter which can be the dominant part in the
overall composition of the hybrid star. A sizable mixed phase exists, in
particular for large values of the bag constant, where the pure quark
core is absent in the hybrid star. For sufficiently large values of the
bag constant, the mixed phase as well as the pure quark phase is not
present in the compact star, which is then an ordinary neutron star
without any quark matter component in its core. The right plot shows the
case, when the quarks have a quasiparticle effective mass due to
interactions via gluon exchange which is parameterised by the coupling
constant $g$. The onset of the mixed phase and the pure quark phase is
shifted to lower values of the bag constant. Note, that the composition
of a compact star with quark matter is highly sensitive to the choice of
the bag constant. Anything between a pure quark star and an ordinary
compact star of hadronic matter only is possible due to our present
limited knowledge about the properties of quark matter at extreme
densities.

\begin{figure}
\centerline{
\includegraphics[width=\textwidth]{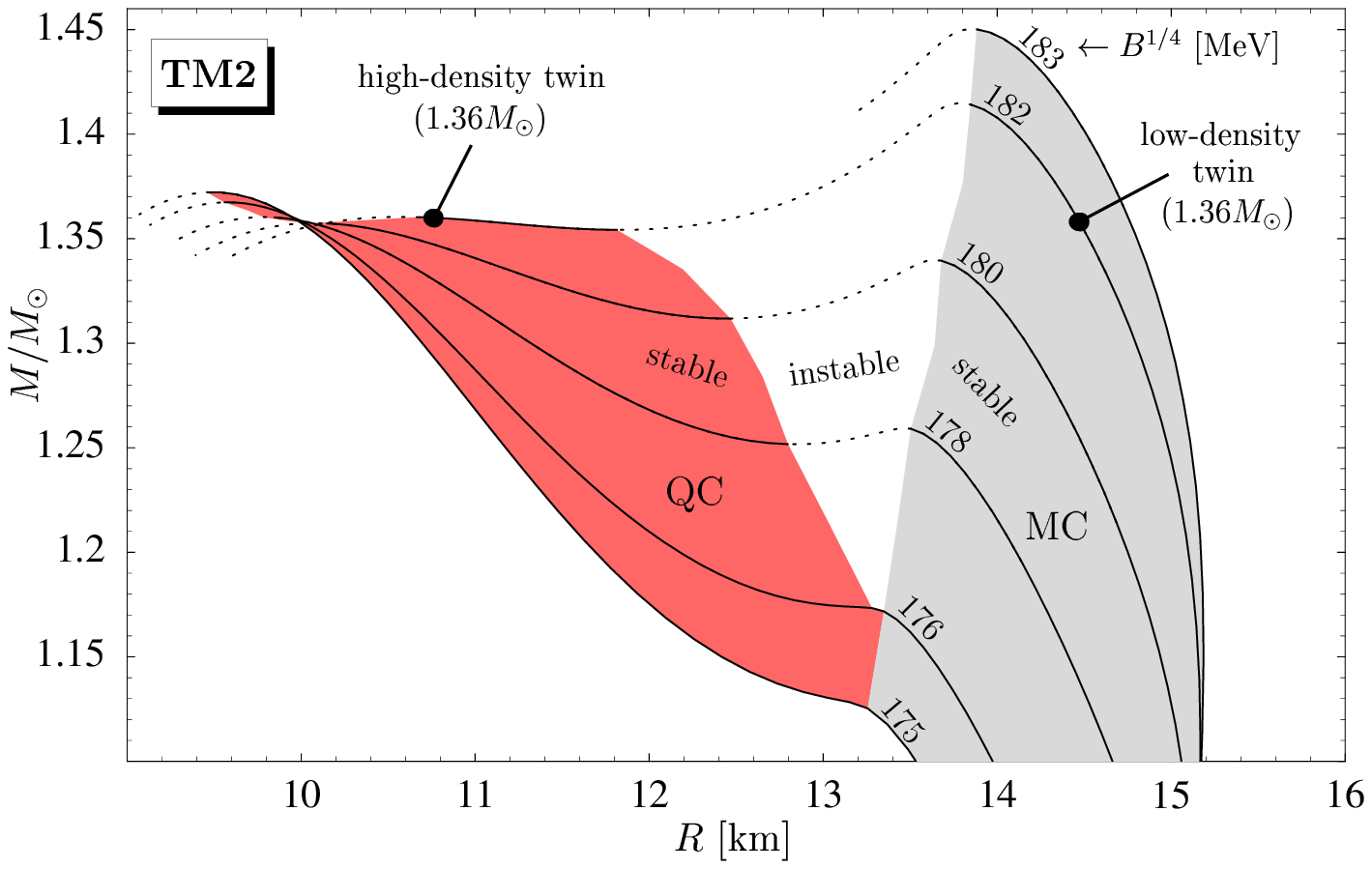}}
\caption{The mass-radius relation of hybrid stars, compact stars with
  hadronic and quark matter (taken from \cite{Schertler:2000xq}).
  Configurations exists where the core consists just of the mixed phase
  (mixed core: MC) or where the core contains pure quark matter (quark
  core: QC). In the latter case, the compact star with a quark core can
  constitute a third family of compact stars with similar masses but
  smaller radii than the mixed core counterparts.}
\label{fig:mr_hybrid}
\end{figure}

Finally, the mass-radius diagram for hybrid stars is shown in
Fig.~\ref{fig:mr_hybrid} for different values of the bag constant.  For
large radii, a mixed phase is formed in the core of the compact star.
The mass is decreasing for lower values of the bag constant (note the
small differences in the values for the bag constant). For some values
of the bag constant, the mass decreases as a function of the central
energy density (for smaller radii) after reaching a maximum mass. This
region is unstable with respect to radial oscillations. However, at even
larger central energy densities (smaller radii), the mass starts to
increase again and a new sequence of stable solutions appear. The
compact stars corresponding to these new stable sequence constitute a
third family of compact stars, besides white dwarfs and ordinary neutron
stars. They are stabilised by the presence of a pure quark phase which
has a sufficiently stiff equation of state so as to support very dense
compact star configurations. It is possible to have two compact stars
with identical masses but with different radii, so-called compact star
twins \cite{Glendenning:1998ag,Schertler:2000xq}.

In recent years, the properties of QCD at high densities have been
explored in connection with the phenomenon of colour-superconductivity
\cite{Alford:1997zt,Rapp:1997zu}, for reviews see
e.g.~\cite{Rischke:2003mt,Alford:2007xm}. A rich phase structure exists
in the QCD phase diagram at high densities and low temperatures right in
the regime of compact star physics. In particular, several phase
transitions might be present in compact star matter, so that the quark
matter core of hybrid stars might contain more than one phase
transition. This research field is rapidly evolving and is poised for new
discoveries for the properties of compact stars with quark matter.  We
refer the interested reader to the above mentioned review articles for
further reading.

\section{Summary}

Compact stars can have a rich structure with new and exotic phases being
present in their cores. Hyperons, heavy baryons with strangeness, can
appear in neutron star matter. As a new fermionic degree of freedom,
hyperons lower the overall pressure so that the maximum mass of a
compact star with hyperons is substantially smaller than the one of a
neutron star consisting of nucleons and leptons only. Also, negatively
charged Bose-Einstein condensate could be formed via kaon condensation
at high densities. As the Bose-Einstein condensate does not give a
direct contribution to the pressure of the system, the stable sequence
of compact stars in the mass-radius diagram simply stops as soon as the
kaon condensed phase appears or extremely dense compact stars are
generated with unusually small radii. Finally, at high densities quark
matter can be present. Depending on the onset of the phase transition
and the strength of the first-order phase transition from the hadronic
to the quark matter phase, a third family of compact stars is found in
the mass-radius diagram. The compact stars of this third family have a
core of pure quark matter and smaller radii than ordinary hybrid stars.
The physics of dense quark matter in compact star opens exciting new
perspectives in exploring the phase diagram of QCD at high densities.
The new research facility FAIR at GSI Darmstadt will explore this
fascinating regime of strong interaction physics in the near future with
relativistic heavy-ion collisions with bombarding energies tuned so as
to create compressed baryonic matter at highest baryon densities. 


\bibliography{literat,all}
\bibliographystyle{revtex}

\end{document}